\documentclass[twocolumn,aps,prb,showpacs,superscriptaddress]{revtex4-1}

\usepackage{amsfonts}
\usepackage{amssymb}
\usepackage{graphicx}
\usepackage{float}
\usepackage{amsmath}
\usepackage{lipsum}

\begin{document}

\title{Signature of nonequilibrium quantum phase transition in the long time
average of Loschmidt echo}
\author{Bozhen Zhou}
\affiliation{Beijing National Laboratory for Condensed Matter Physics, Institute of
Physics, Chinese Academy of Sciences, Beijing 100190, China}
\affiliation{School of Physical Sciences, University of Chinese Academy of Sciences,
Beijing, 100049, China}
\author{Chao Yang}
\affiliation{Beijing National Laboratory for Condensed Matter Physics, Institute of
Physics, Chinese Academy of Sciences, Beijing 100190, China}
\affiliation{School of Physical Sciences, University of Chinese Academy of Sciences,
Beijing, 100049, China}
\author{Shu Chen}
\thanks{Corresponding author: schen@iphy.ac.cn}
\affiliation{Beijing National Laboratory for Condensed Matter Physics, Institute of
Physics, Chinese Academy of Sciences, Beijing 100190, China}
\affiliation{School of Physical Sciences, University of Chinese Academy of Sciences,
Beijing, 100049, China}
\affiliation{Yangtze River Delta Physics Research Center, Liyang, Jiangsu 213300,
China}
\date{ \today }

\begin{abstract}
We unveil the role of the long time average of Loschmidt echo in the
characterization of nonequilibrium quantum phase transitions by studying
sudden quench processes across quantum phase transitions in various quantum
systems. While the dynamical quantum phase transitions are characterized by
the emergence of a series of zero points at critical times during time
evolution, we demonstrate that nonequilibrium quantum phase transitions can
be identified by nonanalyticities in the long time average of Loschmidt
echo. The nonanalytic behaviours are illustrated by a sharp change in the
long time average of Loschmidt echo or the corresponding rate function or
the emergence of divergence in the second derivative of rate function when
the driving quench parameter crosses the phase transition points. The
connection between the second derivative of rate function and fidelity
susceptibility is also discussed.
\end{abstract}

\maketitle


\section{INTRODUCTION}

Over the past decades, quantum phase transitions (QPTs) have been attracted
considerable attention in condense matter physics \cite{Sach}. Contrary to
the classical phase transitions driven by the temperature, QPTs occur at
absolute zero temperature due to quantum fluctuations and are driven by
physical parameters. According to Landau's criteria, QPTs are characterized
by singularities of the ground-state (GS) energy and a nth-order QPT is
defined by discontinuities in the nth derivative of the energy. In recent
years, some new approaches in quantum-information sciences shed light on the
QPTs \cite{AO2002Nature,TJO2002PRA,GV2003PRL,LA2008RMP} and unveil the role
of GS wavefunction in the characterization of QPTs \cite%
{QuanHT,Zanardi,GuSJ,PZ2006PRE}. One of the useful concepts is the GS
fidelity, which is found to exhibit an abrupt drop at the phase transition
point and can be applied to identify a QPT \cite%
{PZ2006PRE,SC2007PRE,ZHQ2008JPAMT41,SC2008PRA,Zanardi,GuSJ}.

Meanwhile, QPTs far from equilibrium systems have extended our understanding
of phase transitions and universality greatly \cite%
{TP2008PRL,TP2011PRL,Werner,Altman,MH2013PRL,Sciolla,Diehl,Diehl-2008,Calabrese,Polkovnikov}%
. By a sudden change of the Hamiltonian, a quantum quench process can push
the initial quantum system out of equilibrium, which permits us to study the
quench dynamics of the nonequilibrium system. More recently, many
researchers concentrated on critical phenomena presented in quench dynamics,
which are termed dynamical quantum phase transitions (DQPTs) \cite%
{MH2013PRL,CK2013PRB,EC2014PRL,FA2014PRB,MM2014PRL,JMH2014PRB,MH2014PRL,MH2015PRL,MS2015PRB,JCB2016PRB,MH2018PRL,YC17PRB,Mera,AA16LTP,Heyl18RPP}.
An important quantity to describe DQPT is Loschmidt echo (LE), which
measures the overlap of an initial quantum state and its time-evolved state
after the quench \cite{Gorin}. For the system with the initial state siting
in ground state of a given Hamiltonian, it is found that the Loschmidt echo
exhibits a series of zero points at critical times $\{t^{\ast}\}$ during
time evolution if the post-quench Hamiltonian and initial Hamiltonian
correspond to different phases. Corresponding to these zero points, the
dynamical free energy density in the thermodynamic limit becomes nonanalytic
as a function time, which is a characteristic feature of DQPT and has been
verified in various systems\cite%
{MH2013PRL,CK2013PRB,EC2014PRL,FA2014PRB,MM2014PRL,JMH2014PRB,MH2014PRL,MH2015PRL,MS2015PRB,JCB2016PRB,MH2018PRL,YC17PRB}.
Review articles about DQPTs can be found in references \cite{AA16LTP,Heyl18RPP}.  The LE has also been found applications in the context of decoherence \cite%
{QuanHT,Rossini}, quantum criticality \cite{Campbell,Jafari,WeiBB},
out-of-equilibrium fluctuations \cite{Venuti,Hamma} and many-body
localization \cite{MBL}. 

Besides the notation of DQPT, the concept of a steady-state transition was
also proposed to describe the nonequilibrium QPT induced by quantum quench.
In this notation, the nonequilibrium QPT is signaled by a nonanalytic change
of physical properties as a function of the quench parameter in the
asymptotic long-time state of the system \cite{Sciolla,Diehl,Diehl-2008}.
Usually, time average of order parameter was used to characterize
nonequilibrium criticality \cite{Silva2018}. The connection between the DQPT
and steady-state transition was addressed recently \cite%
{Silva2018,Halimeh,WangPei2018}. Although the notation of LE plays a
particularly important role in the characterization of DQPT, its connection
to the nonequilibrium QPT is not well understood yet.

In this work, we shall explore the role of long time average of LE in the
characterization of nonequilibrium QPT. The long time average of LE is
independent of time and conveys information of overlap of the initial state
and eigenstats of the post-quench Hamiltonian. By studying several typical
models which exhibit nonequilibrium QPTs, we demonstrate that the long time
average of LE or closely related quantities display nonanalytic behavior
when the quench parameter crosses a quantum phase transition point,
suggesting that the nonanalytic change of long time average of LE can give
signature of nonequilibrium QPT. For the specific case with the pre-quench
and post-quench parameters being very close, we find that there exists an
equivalent relation between the second derivative of rate function of long
time average of LE and fidelity susceptibility, which indicates the
existence of divergence in the second derivative of the rate function at the
phase transition point.

\section{Long time average of Loschmidt echo and quench dynamics}

\subsection{long time average of Loschmidt echo}

Without loss generality, we consider a general Hamiltonian undergoing a QPT
described by $H(\lambda)$, where $\lambda$ is a control parameter which
drives the QPT. Suppose the system is initially prepared as the ground state
of the Hamiltonian $H(\lambda_i)$, we investigate the quench dynamics by
suddenly changing the driving parameter to $\lambda_f$, i.e., the sudden
quench process is realized by a sudden change of the control parameter
\begin{equation*}
\lambda(t) = \lambda_i \theta(-t) + \lambda_f \theta(t),
\end{equation*}
where $\lambda_i$ represents the control parameter in the initial
(pre-quench) Hamiltonian, $\lambda_f$ the control parameter in the final
(post-quench) Hamiltonian, and $\theta(t)$ is the Heaviside step function.
Before studying concrete models, we shall briefly introduce the notation of
LE and give the expression of long time average of LE. Given an initial
quantum state $|\Psi(0)\rangle $, the Loschmidt amplitude is defined as
\begin{equation}
\mathcal{G}(t)=\langle \Psi(0)|\Psi(t)\rangle = \langle
\Psi(0)|e^{-iH(\lambda_f)t}|\Psi(0)\rangle,
\end{equation}
which represents the overlap between the initial state and the time-evolved
state after the quantum quench, and the Loschmidt echo is given by
\begin{equation}
\mathcal{L}(t)=|\mathcal{G}(t)|^{2}
\end{equation}
which is the probability associated with Loschmidt amplitude. Traditionally,
the ground state of the initial Hamiltonian is chosen as the initial state,
and the Loschmidt echo could be interpreted as the return probability of the
ground state during time evolution.

In this work, we mainly focus on the long time average of LE
\begin{equation}
\overline{\mathcal{L}}\equiv \lim_{\tau \rightarrow \infty }\frac{1}{\tau }%
\int_{0}^{\tau }|\mathcal{G}(t)|^{2}\mathtt{d}t.
\end{equation}%
By using
\begin{equation*}
|\Psi (t)\rangle =e^{-iH(\lambda _{f})t}|\Psi (0)\rangle
=\sum_{n}e^{-iE_{n}t}|\psi _{n}\rangle \langle \psi _{n}|\Psi (0)\rangle ,
\end{equation*}%
it follows
\begin{equation}
\overline{\mathcal{L}}(\lambda _{f})=\sum_{n}\left\vert \langle \psi
_{n}(\lambda _{f})|\Psi (0)\rangle \right\vert ^{4},
\end{equation}%
where the evolved state is expanded by the normalized eigenstates of $%
H(\lambda _{f})$ denoted by $|\psi _{n}\rangle $ with eigenenergy $E_{n}$.
It can be found that the long time average of LE has a similar form of
inverse participation ratio \cite{AD2013EPL}, which gives the distribution
information of the initial state in the Hilbert space of the post-quench
Hamiltonian. In order to study critical properties of many-particle systems,
we introduce the rate function of $\overline{\mathcal{L}}$ ,
\begin{equation}
\eta (\lambda _{f})=-\frac{1}{L}\log \overline{\mathcal{L}}(\lambda _{f}),
\end{equation}%
which is defined as the logarithm of $\overline{\mathcal{L}}$ divided by the
system size $L$ and is an intensive quantity in the thermodynamic limit.
When $\lambda _{f}$ approaches a critical point $\lambda _{c}$, $\eta
(\lambda _{f})$ shall exhibit nonanalytic behaviours, which can be viewed as
a characteristic signature of nonequilibrium quantum phase transition.

\subsection{Quantum quench in the Aubry-Andr\'{e} model}

We first consider the Aubry-Andr\'{e} (AA) model with Hamiltonian%
\begin{equation}
H=-J\sum_{j=1}^{L-1}(c_{j}^{\dagger }c_{j+1}+h.c.)+\Delta \sum_{j=1}^{L}\cos
(2\pi \alpha j)c_{j}^{\dagger }c_{j},  \label{HAA}
\end{equation}%
where $c_{j}^{\dagger }(c_{j})$ denotes the creation (annihilation) operator
of fermions at site $j$ ($j=1,\cdots ,L$ with $L$ the total number of
lattice sites), $J$ is hopping amplitude and $\Delta $ is the strength of
the incommensurate potential. Here $\alpha $ is an irrational number and we
fix $\alpha =\frac{\sqrt{5}-1}{2}$ for convenience. The incommensurate
potential strength $\Delta $ drives the system undergoing a
delocalization-localization transition at a critical point $\Delta /J=2$.
When $\Delta /J<2$, all the eigenstates are extended, but localized as $%
\Delta /J>2$. \cite{AA,YC17PRB}.

\begin{figure}[tbph]
\includegraphics[width=9cm]{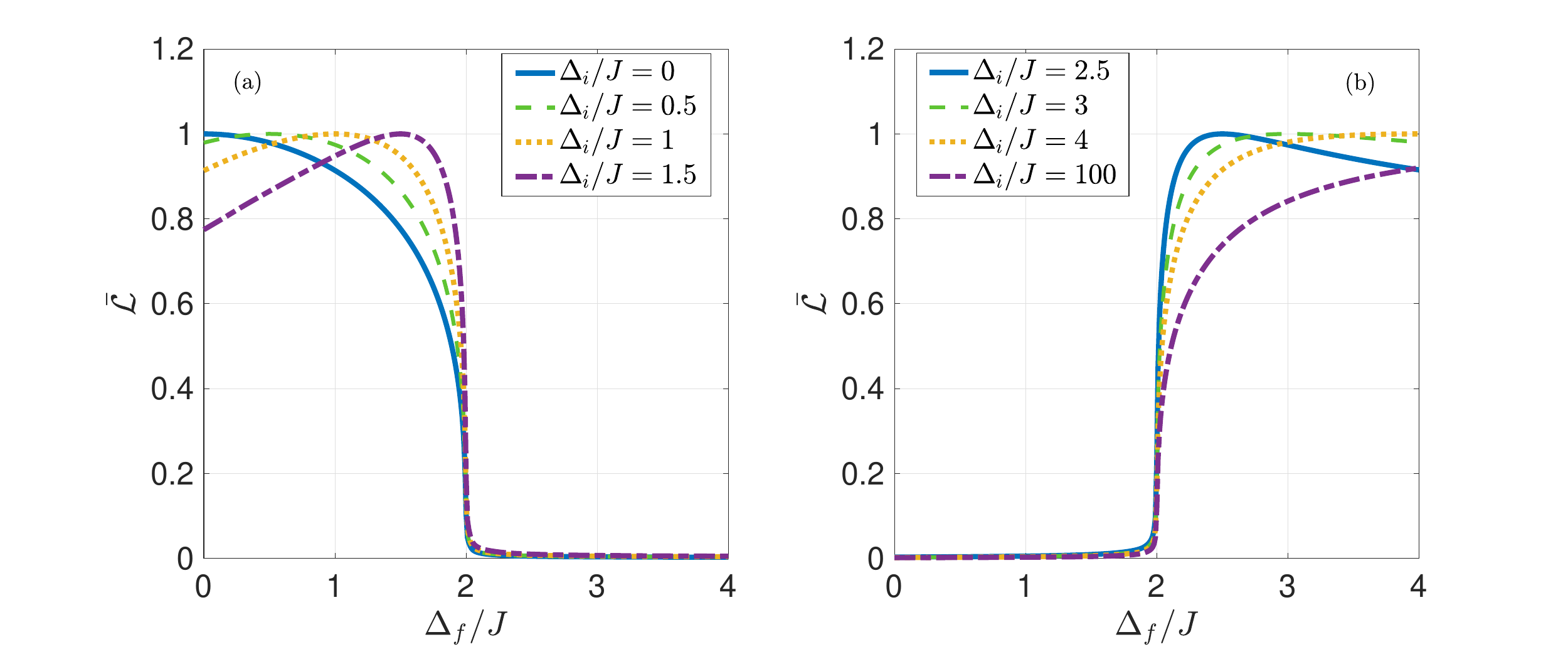}
\caption{(Color online) The behavior of the long time average of LE of AA
model with the total number of lattice sites $L=1000$. The strength of
incommensurate potential in the initial Hamiltonian is (a)  $\Delta_{i}/J=0,\,0.5,\,1,\,1.5$, (b) $\Delta_{i}/J=2.5,\,3,\,4,\,100$, respectively. }
\label{fig1}
\end{figure}

Now we consider the quench process described by the sudden change of the
incommensurate potential strength $\Delta (t)=\Delta _{i}\theta (-t)+\Delta
_{f}\theta (t)$, i.e., we prepare the initial state of system in the ground
state of Hamiltonian $H(\Delta _{i})$, and then suddenly quench to
Hamiltonian $H(\Delta _{f})$ at $t=0$. The DQPT in the AA model has been
studied in Ref.{\cite{YC17PRB}}. It was shown that the LE supports a series
of zero points at critical times if $H(\Delta _{i})$ and $H(\Delta _{f})$
are in different phases. In Fig. \ref{fig1}, we display the long time
average of LE versus $\Delta _{f}/J$ by fixing $\Delta _{i}/J=0,\,0.5,\,1,\,1.5$ (Fig.\ref{fig1}(a)) and $\Delta _{i}/J=2.5,\,3,\,4,\,100$ (Fig.\ref{fig1}(b)), respectively. For
both cases, it is shown that $\overline{\mathcal{L}}$ has an obvious change
around the transition point $\Delta _{f}/J=2$. Therefore, the sharp change of $\overline{\mathcal{L}}$ at the
transition point can give us a characteristic signature of nonequilibrium
quantum phase transition.

\subsection{Quantum quench in the quantum Ising model}

Next we consider the transverse field Ising model described by the following
Hamiltonian
\begin{equation}
H=-J\sum_{j=1}^{L-1} \sigma _{j}^{x}\sigma _{j+1}^{x} +
h\sum_{j=1}^{L}\sigma _{j}^{z}\, ,  \label{HIsing}
\end{equation}%
where $\sigma _{j}^{\alpha }$ $,$ ($\alpha =x,y,z,$) are the Pauli matrices,
$j=1,\cdots ,L$ with $L$ the total number of lattice sites, $J$ is
nearest-neighbor spin exchange interaction, and $h$ is the external magnetic
field along the $z$ axis. The transverse field Ising model can be mapped to
spinless fermions by using Jordan-Wigner transformation: $\sigma _{j}^{z} =
2c_{j}^{\dagger }c_{j} -1 $ and $\sigma _{j}^{x}
=\prod_{i<j}(1-2c_{i}^{\dagger}c_{i})(c_{j}+c_{j}^{\dagger })$. In the
fermion representation, we have
\begin{equation}
H=-J\sum_{j=1}^{L-1}\left( c_{j}^{\dagger }c_{j+1}+c_{j}^{\dagger
}c_{j+1}^{\dagger }+h.c.\right) +2h\sum_{j=1}^{L}c_{j}^{\dagger }c_{j} ,
\end{equation}%
where we have discarded the constant $-hL$ which merely shifts the origin of
energy and has no effect on the phase transition. Now, we consider the
periodic boundary condition and use the Fourier transform $c_{j}^{\dag }=%
\frac{1}{\sqrt{L}}\sum_{k}e^{ikj}c^{\dag }(k)$, where $k$ is the wave vector
and $-\pi <k\leqslant \pi $. In the momentum representation, the
Bogoliubov--de Gennes Hamiltonian is given by
\begin{equation}
H_{k}=\left[
\begin{array}{cc}
-J\cos k+h & iJ\sin k \\
-iJ\sin k & J\cos k-h%
\end{array}%
\right] .
\end{equation}

Introducing a unitary transformation $\mathcal{H}_k=UH_kU^{-1}$ with
\begin{equation}
U=\frac{1}{\sqrt{2}}\left[
\begin{array}{cc}
1 & -1 \\
1 & 1%
\end{array}%
\right] ,
\end{equation}%
then the Hamiltonian is transformed as
\begin{equation}
\mathcal{H}_k=\left[
\begin{array}{cc}
0 & V(k) \\
V^{\ast }(k) & 0%
\end{array}%
\right] ,
\end{equation}%
where $V(k)=h-Je^{-ik}$. The two eigenvalues are $E_{\pm }=\pm \sqrt{%
V(k)V^{\ast }(k)}$ and two eigenvectors are $|\psi _{\pm }\rangle =\frac{1}{%
\sqrt{2}}\left( \left( \frac{V(k)}{V^{\ast }(k)}\right) ^{1/4},\pm \left(
\frac{V^{\ast }(k)}{V(k)}\right) ^{1/4}\right) ^{\text{T}}$. There are two
distinct phases which can be characterized by the winding number $\nu $ with
the from
\begin{equation}
\nu =-\frac{1}{2\pi i}\int_{-\pi }^{\pi }V^{-1}(k)\partial _{k}V(k)dk\, ,
\end{equation}%
where the winding number is either $0$ or $1$, depending on the parameters.
It follows that the winding number $\nu =1$ for $|h/J|\leqslant 1$
corresponding to the topological phase, otherwise $\nu =0$ represents the
trivial phase. We focus on the region of $h/J\geqslant 0$ and the phase
transition point is given by $h_{c}/J=1$. 
\begin{figure}[tbph]
\includegraphics[width=9cm]{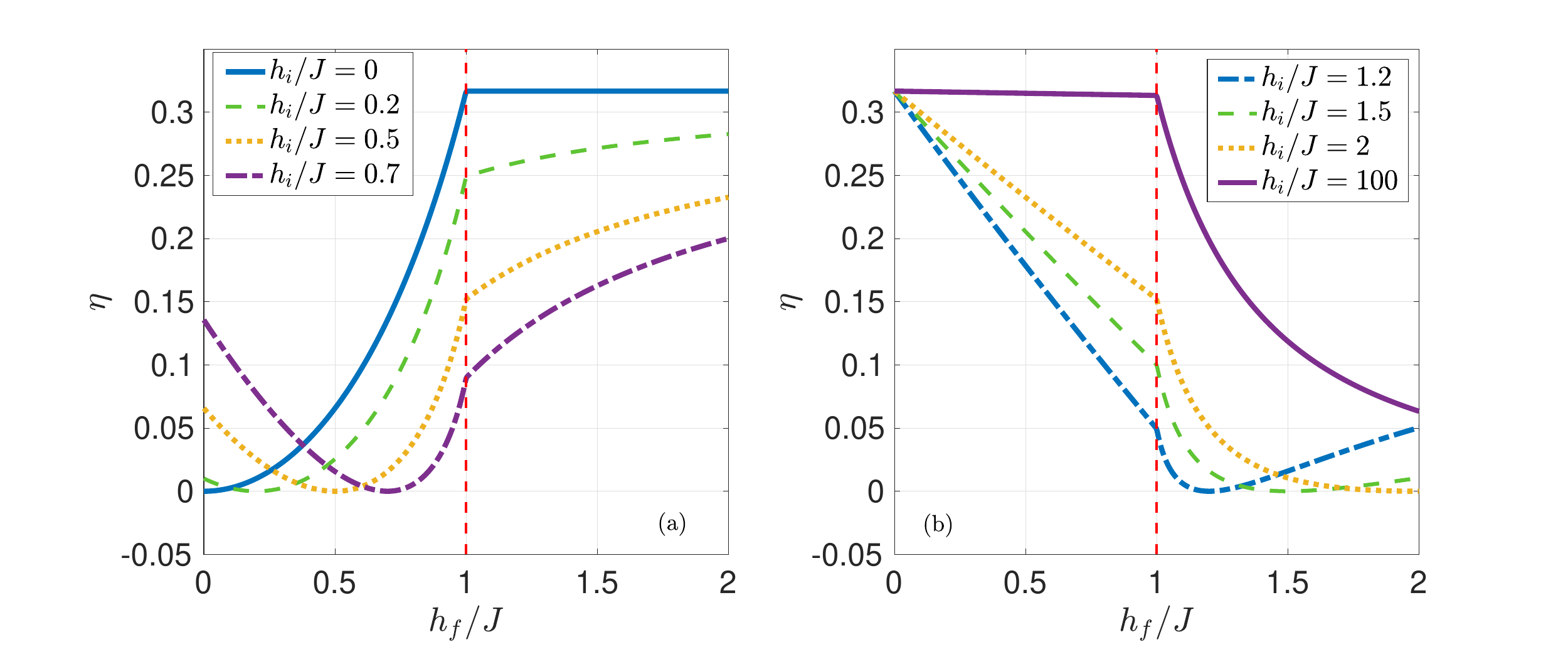}
\caption{(Color online) The behavior of $\protect\eta $ of Ising model with
respect to external magnetic field $h_{f}$ of the post-quench Hamiltonian.
The total number of lattice sites $L=1000$. The red dashed vertical line in figures
guides the value of the phase transition point $h_{c}/J=1$. The external
magnetic field along the $x$ axis in the intital Hamiltonian is (a) $h_{i}/J=0,\,0.2,\,0.5,\,0.7$, (b) $h_{i}/J=1.2,\,1.5,\,2,\,100$, respectively. }
\label{LLEIsing}
\end{figure}

Now we consider the quench process with $h(t)=h_{i}\theta (-t)+h_{f}\theta
(t)$. We prepare the ground state of quantum Ising model $H(h_{i})$ in
fermion representation as the initial state. For convenience, we can
calculate the rate function of the long time average of LE which has the
form {\small
\begin{eqnarray}
\eta (h_{f}) &=&-\frac{1}{L}\log \left[ \sum_{\alpha _{1}\cdots \alpha
_{N}=\pm }\left\vert \vphantom{\sum_a}\left( \langle \phi _{\alpha
_{1}}(k_{1})|\otimes \cdots \otimes \langle \phi _{\alpha
_{N}}(k_{N})|\right) \right. \right.  \notag \\
&&\left. \left. \times \left( |\psi _{-}(k_{1})\rangle \otimes \cdots
\otimes |\psi _{-}(k_{N})\rangle \right) \vphantom{\sum_a}\right\vert
^{4}\mathstrut \vphantom{\sum_{\alpha_1}}\right].
\end{eqnarray}
} In the limit of $L\rightarrow\infty$, the momentum $k$ distributes
continuously and we can get
\begin{eqnarray}
\eta (h_{f}) = -\frac{1}{2\pi }\int_{-\pi }^{\pi }\text{d}k\left[ \log
\sum_{\alpha =\pm }\left\vert \vphantom{\sum}\langle \phi _{\alpha }(k)|\psi
_{-}(k)\rangle \right\vert ^{4}\right] ,
\end{eqnarray}
where $|\psi _{-}(k)\rangle $ is the ground state wavefunction of the
initial Hamiltonian $\mathcal{H}_{k}(h_{i})$. Then the time evolution is
governed by the final Hamiltonian $\mathcal{H}_{k}(h_{f})$ with two
eigenvalues $E_{\pm }(h_{f})$ and two corresponding wavefunctions are $|\phi
_{\alpha }(k)\rangle$ ($\alpha =\pm$). Substituting the concrete form of $%
V(k)$ in $|\psi _{-}(k)\rangle $ and $|\phi _{\alpha }(k)\rangle $. Then $%
\eta$ can be written as
\begin{equation}
\eta =-\frac{1}{2\pi }\int_{-\pi }^{\pi }\text{d}k\left[ \log \frac{1+\cos
^{2}\theta }{2}\right]  \label{LLEI}
\end{equation}%
with
\begin{equation*}
\theta =\arctan \frac{\left( h_{f}/J-h_{i}/J\right) \sin k}{%
1+h_{i}h_{f}/J^2-\left( h_{i}/J+h_{f}/J\right) \cos k},
\end{equation*}
where $h_{i}/J$ and $h_{f}/J$ are external magnetic field along the $x$ axis
in the initial and final Hamiltonian, respectively.

We numerically calculate Eq.(\ref{LLEI}) and show several results with different initial Hamiltonian in Fig.\ref%
{LLEIsing}. For Fig.\ref{LLEIsing}(a), taking the initial state prepared in the
phase with $h_{i}/J=0$ and $\nu =1$ (solid line) as an example, we can see that $\eta $ grows from $0$
to the value approximately equal to $0.315$ as $h_{f}/J$ increases from $0$
to the critical point $h_{c}/J=1$, where $\eta =0$ means the final state is
the same as the initial state with $h_{i}=h_{f}$. When the parameter $h_{f}/J
$ crosses the critical point, the final Hamiltonian enters into the trivial
phase with $\nu =0$, and $\eta $ keeps approximately to be a constant with
the increasing of $h_{f}/J$. In Fig.\ref{LLEIsing}(b), taking the
initial Hamiltonian in the trivial phase with $h_{i}/J=100$ and $\nu =0$ (solid line) as an example,
and continuously change the parameter $h_{f}/J$ of final Hamiltonian from $0$
to $2$. It can be seen that $\eta $ remains a constant approximately equal
to $0.315$ when the final Hamiltonian is in the topological phase with $%
h_{f}\leqslant h_{c}$. After crossing the critical point $h_{c}/J$ with $%
h_{f}>h_{c}$, $\eta $ begins to decrease with the increase of $h_{f}$ and
shall reach the minimum value $0$ at $h_{i}=h_{f}$.  Generally, we can see the nonanalyticity of $\eta $ emerges as long as $h_{f}$ crosses
the critical point and is independent of the choice of the initial state.
Nonequilibrium QPT is characterized by the nonanalytic behavior of $\eta $
at the critical point.

\subsection{Quantum quench in the Haldane model}

\begin{figure}[tbph]
\includegraphics[width=8.5cm]{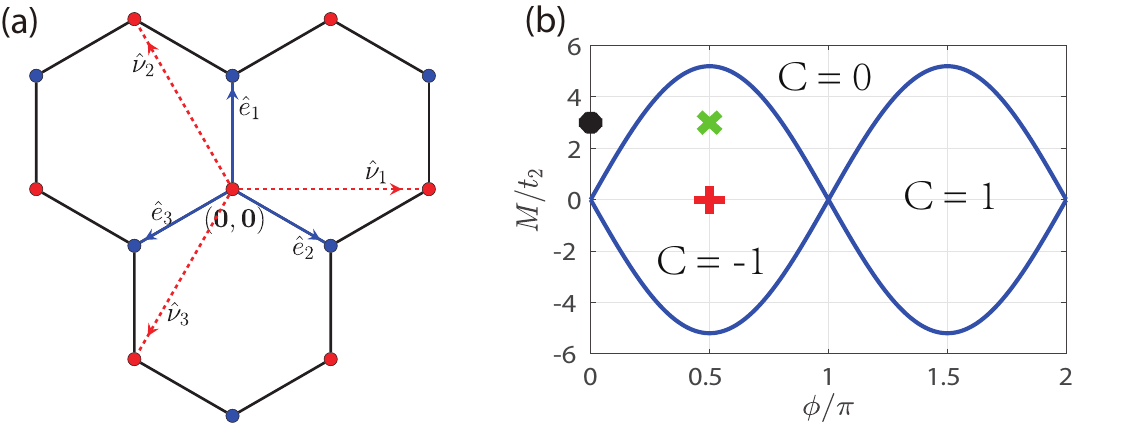}
\caption{(Color online) (a) Illustration of the honeycomb lattice. Red and
blue circles represent two sublattices (A and B). Three blue lines with
arrow denote three nearest-neighbor displacements of A, and three red dashed
lines with arrow denote three distinct next-nearest-neighbor displacements
of A. (b) Phase diagram of Haldane model. Red plus sign marks the initial
state for Fig.4. Black dot marks the initial state for Fig.5(a) and (b) and
green times sign marks the initial state for Fig.5(c) and (d). }
\label{HoneyLattice}
\end{figure}

In this subsection, we investigate the Haldane model \cite{Haldane}
described by the following tight-binding Hamiltonian {\small
\begin{equation}
{H=M\sum_{j} \left[ c_{A,\vec{r}_{j}}^{\dag }c_{A,\vec{r}_{j}}-c_{B,\vec{r}%
_{j}}^{\dag }c_{B,\vec{r}_{j}}\right] +H_{\text{NN}}+H_{\text{NNN}}} ,
\label{HaldaneHN}
\end{equation}%
}
with
\begin{eqnarray}
H_{\text{NN}} &=&-t_{1}\sum_{j} \left[ c_{A,\vec{r}_{j}}^{\dag }c_{B,\vec{r}%
_{j}+\widehat{e}_{1}}+c_{A,\vec{r}_{j}}^{\dag }c_{B,\vec{r}_{j}+\widehat{e}%
_{2}}\right.  \notag \\
&&+\left. c_{A,\vec{r}_{j}}^{\dag }c_{B,\vec{r}_{j}+\widehat{e}_{3}}+h.c.%
\right] ,  \label{HaldaneHNN}
\end{eqnarray}
and
\begin{eqnarray}
H_{\text{NNN}} &=&-t_{2}e^{i\phi }\sum_{j} \left[ c_{A,\vec{r}_{j}}^{\dag
}c_{A,\vec{r}_{j}+\widehat{\nu }_{1}}+c_{A,\vec{r}_{j}}^{\dag }c_{A,\vec{r}%
_{j}+\widehat{\nu }_{2}}\right.  \notag \\
&&+\left. c_{A,\vec{r}_{j}}^{\dag }c_{A,\vec{r}_{j}+\widehat{\nu }%
_{3}}+(A\rightarrow B)+h.c.\right] ,  \label{HaldaneHNNN}
\end{eqnarray}%
where the on-site energy is $M$ on $A$ sites and $-M$ on $B$ sites, $H_{%
\text{NN}}$ denotes the Hamiltonian with nearest-neighbor (NN) hopping
amplitude $t_{1}$, and $H_{\text{NNN}}$ the Hamiltonian with
next-nearest-neighbor (NNN) hopping amplitude $t_{2}$ and phase difference $%
\phi $, Here $c_{\alpha,\vec{r}_{j}}^{\dagger }(c_{\vec{r}_{j}})$ denotes
the creation (annihilation) operator of fermions at the sublattice $%
\alpha=A, B$ of site $\vec{r}_{j}$. The summation is defined on a
two-dimensional honeycomb lattice. The illustration of honeycomb lattice of
Haldane model is shown in Fig.\ref{HoneyLattice}(a), where $\hat{e}_1 =(0,a)$%
, $\hat{e}_2 =(-\frac{\sqrt{3}}{2} a,-\frac{1}{2} a)$ and $\hat{e}_3 =(\frac{%
\sqrt{3}}{2} a,-\frac{1}{2} a)$ are the displacements from a $A$ site
located at $\vec{r}_{j} $ to its three nearest-neighbor $B$ sites, and $\hat{%
\nu}_1 =(\sqrt{3}a,0)$, $\hat{\nu}_2=(-\frac{\sqrt{3}}{2}a,\frac{3}{2} a)$
and $\hat{\nu}_3 =(-\frac{\sqrt{3}}{2} a,-\frac{3}{2} a)$ are the
displacements from a $A$ site located at $\vec{r}_{j}$ to its three distinct
next-nearest-neighbor $A$ sites. Here $a$ is lattice constant and we shall
fix $a=1$.

By taking the periodic boundary condition along the $x$-axis and $y$-axis
direction,
the Hamiltonian in momentum space can be written as
\begin{widetext}
\begin{equation}
H_{k}=\left[
\begin{array}{cc}
M-2t_{2}\sum_{j=1}^{3}\left[ \cos \phi \cos (\mathbf{k\cdot }\widehat{\nu }%
_{j})-\sin \phi \sin (\mathbf{k\cdot }\widehat{\nu }_{j})\right]  &
-t_{1}\sum_{j=1}^{3}\left[ \cos (\mathbf{k\cdot }\widehat{e}_{j})-i\sin (%
\mathbf{k\cdot }\widehat{e}_{j})\right]  \\
-t_{1}\sum_{j=1}^{3}\left[ \cos (\mathbf{k\cdot }\widehat{e}_{j})+i\sin (%
\mathbf{k\cdot }\widehat{e}_{j})\right]  & -M-2t_{2}\sum_{j=1}^{3}\left[
\cos \phi \cos (\mathbf{k\cdot }\widehat{\nu }_{j})-\sin \phi \sin (\mathbf{%
k\cdot }\widehat{\nu }_{j})\right]
\end{array}%
\right],
\end{equation}
\end{widetext}
where $\mathbf{k}$ is the wavevector in the first Brillouin zone (FBZ). The
topologically different phases of Haldane model can be characterized by
Chern number with the form \cite{Haldane}
\begin{equation}
C=\frac{1}{2\pi }\int_{\text{FBZ}}\Omega _{n}(\mathbf{k)}\text{d}\mathbf{k}
\, ,
\end{equation}%
where $\Omega _{n}(\mathbf{k)}$ is the Berry curvature of the $n$-th band
with the Berry connection $\mathcal{A}_{n}(\mathbf{k})=-i\langle \phi _{n}(%
\mathbf{k)}|\nabla _{\mathbf{k}}|\phi _{n}(\mathbf{k)}\rangle $. The phase
diagram in the $(\phi ,M)$ plane is shown in Fig.\ref{HoneyLattice}(b).
While the regime of $C=0$ represents the topologically trivial phase,
regimes with $C=\pm 1$ represent topological phases.

\begin{figure}[tbph]
\includegraphics[width=9cm]{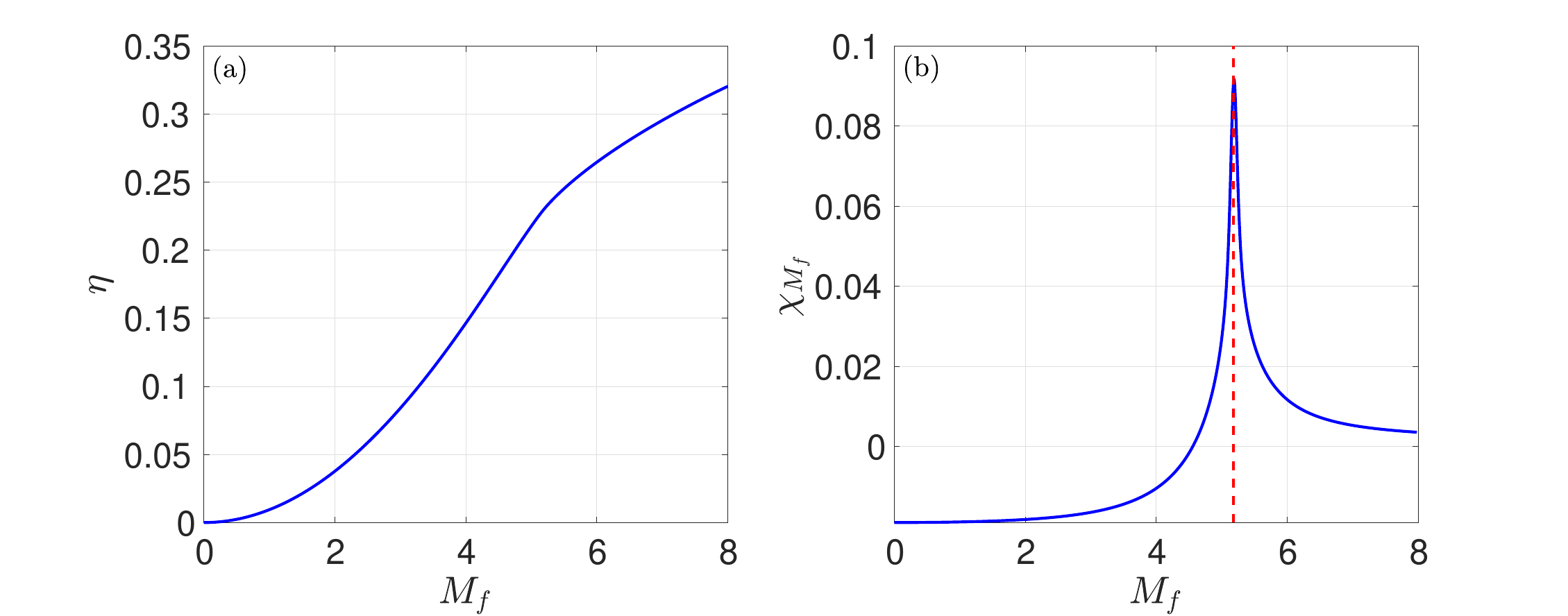}
\caption{(Color online) (a) The behavior of $\protect\eta $ of Haldane model
with respect to on-site energy $M_{f}$ of the final Hamiltonian. (b) The
second derivative of $\protect\eta$ versus $M_{f}$. The red dashed line in
figure guides the value of the topological phase transition point. We have
taken $t_{1}=4,t_{2}=1,$ and $\protect\phi = \protect\pi /2$. The total
number of lattice sites is $L=30082$. The on-site energy of the initial
Hamiltonian is $M_{i}=0$.}
\label{HAD_M}
\end{figure}

Now we consider the quench process solely driven by either the parameter $M$
or the phase difference $\phi $, i.e., the sudden quench described by $%
M(t)=M_{i}\theta (-t)+M_{f}\theta (t)$ or $\phi (t)=\phi _{i}\theta
(-t)+\phi _{f}\theta (t)$. Similarly, we calculate the rate function of the
long time average of LE, which takes the following form:%
\begin{equation}
\eta =-\frac{1}{L}\sum_{j}\left[ \log \sum_{\alpha_j =\pm }\left\vert %
\vphantom{\sum}\langle \phi _{\alpha_j }(\mathbf{k}_j)|\psi _{-}(\mathbf{k}%
_j)\rangle \right\vert ^{4}\right] ,
\end{equation}%
where $L$ is the total number of lattice sites. As the system approaches the
thermodynamic limit, the rate function of the long time average of LE takes
the continuous form:%
\begin{equation}
\eta=- \frac{1}{S_{k}}\int_{\text{FBZ}}\text{d}\mathbf{k}\left[ \log
\sum_{\alpha =\pm }\left\vert \vphantom{\sum}\langle \phi _{\alpha }(\mathbf{%
k})|\psi _{-}(\mathbf{k})\rangle \right\vert ^{4}\right],
\end{equation}%
where $S_{k}$ is the area of FBZ, $|\psi _{-}(\mathbf{k})\rangle $ is the
ground state wavefunction of the initial Hamiltonian in momentum space, and $%
|\phi _{\pm }(\mathbf{k})\rangle $ are wavefunctions of the final
Hamiltonian. By preparing the initial state in the topologically nontrivial
phase with $M_{i}=0$, $\phi =0.5\pi $ and $C=-1$ corresponding to the red
plus sign marked in Fig.\ref{HoneyLattice}(b), we study the quench dynamics
driven by the final Hamiltonian with different $M_{f}$. The behaviour of $%
\eta $ versus $M_{f}$ is illustrated in Fig.\ref{HAD_M}(a). As $\eta $ grows
from $0 $ with increasing $M_{f}$ from $0$ to $8$, no an obvious change is
observed when $M_{f}$ crosses the phase transition point. Nevertheless, we
can define the quantity $\chi _{\lambda _{f}}$ which is equal to the minus
of the second derivative of $\eta $ with respect to the post-quench
parameter $\lambda _{f}$:
\begin{equation}
\chi _{\lambda _{f}}=-\frac{\partial ^{2}\eta }{\partial \lambda _{f}^{2}}\,
.
\end{equation}%
We find that $\chi _{M_{f}}$ exhibits discontinuity with an obvious peak
around $M_{f}=3\sqrt{3}$ as shown in Fig.\ref{HAD_M}(b). The value of $M_{f}$
at discontinuous point of $\chi _{M_{f}}$ is exactly equal to the value of
topological phase transition point calculated by Chern number.

\begin{figure}[tbph]
\includegraphics[width=9cm]{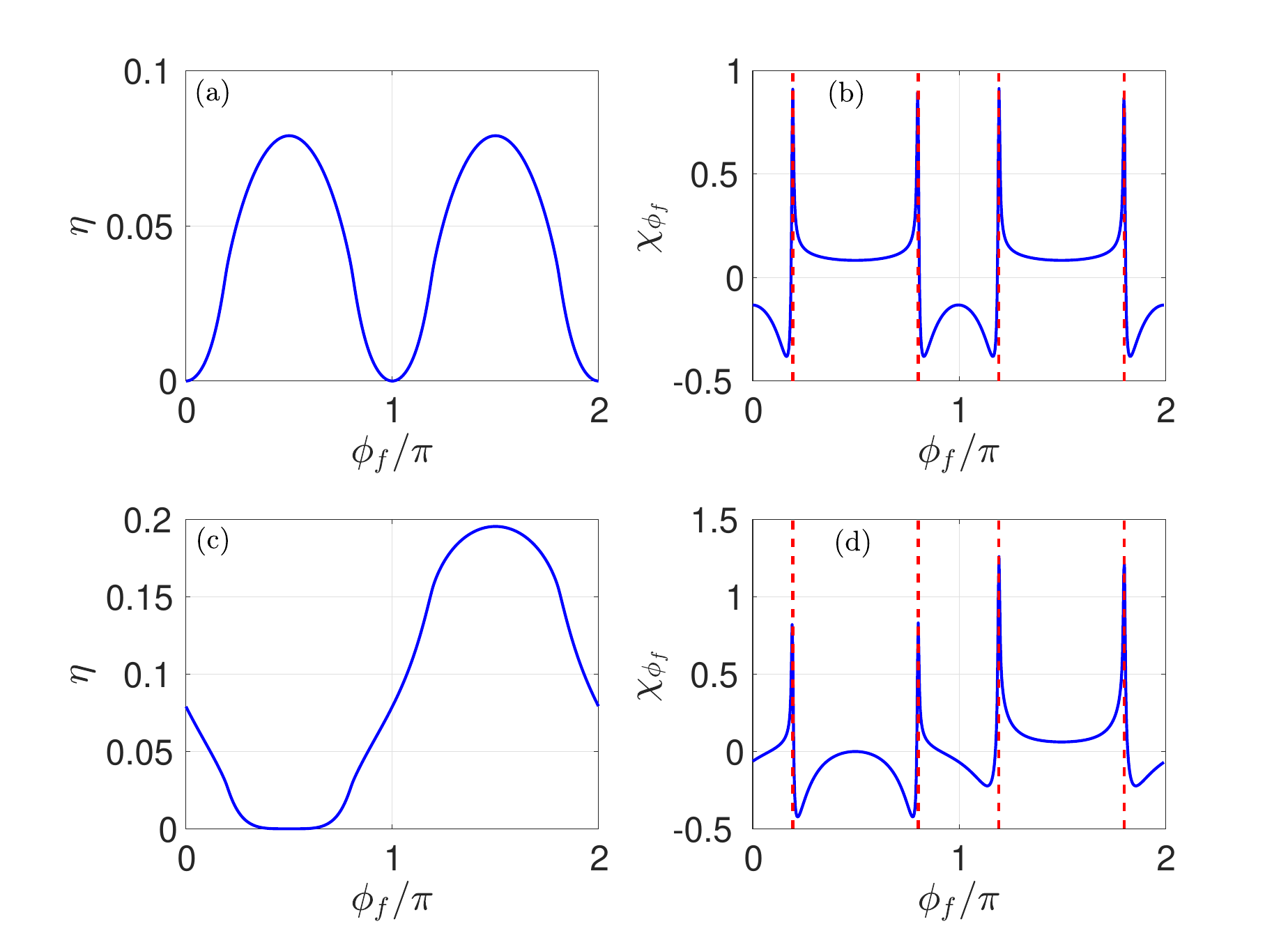}
\caption{(Color online) (a) The behavior of $\protect\eta $ of Haldane model
with respect to $\protect\phi _{f}$ of the final Hamiltonian. (b) The second
derivative of $\protect\eta $ versus $\protect\phi _{f}$. The red dashed
lines in the figure guide the values of the topological phase transition
points which are calculated by Chern number. We have taken $t_{1}=4$, $%
t_{2}=1$ and $M=3$. The total number of lattice sites is $L=30082$. The
phase difference of initial Hamiltonian is (a), (b) $\protect\phi _{i}=0$,
and (c), (d) $\protect\phi _{i}=\protect\pi /2$. }
\label{HAD_phi}
\end{figure}

Next, we study the quench dynamics driven by the final Hamiltonian with
different $\phi _{f}$. The initial state corresponding to Fig.\ref{HAD_phi}%
(a),(b) is prepared in the topologically trivial phase with $M_i = 3\sqrt{3}%
, \phi= 0$ and $C = 0$ as marked by black dot in Fig.\ref{HoneyLattice}(b),
and the initial state corresponding to Fig.\ref{HAD_phi}(c),(d) is prepared
in the topologically nontrivial phase with $M_i = 3\sqrt{3}, \phi= 0.5\pi$
and $C = -1$ as marked by green times sign in Fig.\ref{HoneyLattice}(b). We
display $\eta $ versus $\phi _{f}$ in Fig.\ref{HAD_phi}(a) and $\chi _{\phi
_{f}}$ versus $\phi _{f}$ in Fig.\ref{HAD_phi}(b). While no obvious
nonanalyticity is found in Fig.\ref{HAD_phi}(a), $\chi _{\phi _{f}}$
exhibits discontinuities with obvious peaks at $\phi _{f}\approx 0.194\pi
,0.802\pi ,1.194\pi $ and $1.802\pi $, corresponding to the phase boundaries
in the phase diagram of Fig.\ref{HoneyLattice}(b). For the initial state
prepared in the topological phase with $\phi _{i}=\pi /2$, we display $\eta $
versus $\phi _{f}$ in Fig.\ref{HAD_phi}(c) and $\chi _{\phi _{f}}$ versus $%
\phi _{f}$ in Fig.\ref{HAD_phi}(d). Similarly, we identify four divergent
points at $\phi _{f}\approx 0.194\pi ,0.802\pi ,1.194\pi $ and $1.802\pi $
in Fig.\ref{HAD_phi}(d), whose positions are identical to those displayed in
Fig.\ref{HAD_phi}(b).

\begin{figure}[tbph]
\includegraphics[width=8.5cm]{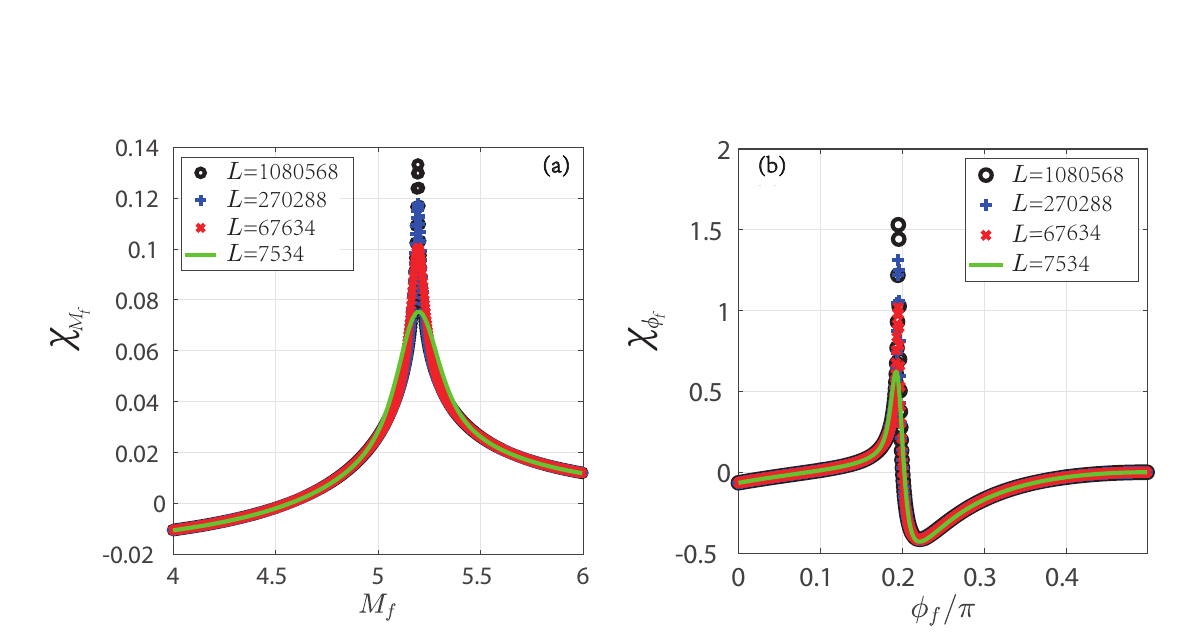}
\caption{(Color online) $\protect\chi _{\protect\lambda_f }$ versus $\protect%
\lambda_f$ for different lattice sizes. The parameters are $t_{1}=4$ and $%
t_{2}=1$. (a) $\protect\phi=\protect\pi/2$ and on-site energy of the initial
Hamiltonian $M_{i}=0$. (b) $M = 3 $ and phase difference of the initial
Hamiltonian $\protect\phi _{i}=\protect\pi /2$ .}
\label{Peak_M_Phi}
\end{figure}

Our results indicate that $\chi _{\lambda _{f}}$ exhibits singular behavior
with the emergence of an obvious peak when the driving parameter crosses the
phase transition point, regardless of our choice of initial state. In Fig.%
\ref{Peak_M_Phi}, we display $\chi _{\lambda _{f}}$ for different lattice
sizes. Despite no real divergence for the finite size system, it is shown
that the height of peak increasing with the lattice size, suggesting the
existence of divergence in the thermodynamic limit.

\subsection{Relation to fidelity susceptibility}

We consider the limiting case that the driving parameters before and after
sudden quench are very close, i.e., $\lambda _{i}=\lambda $ and $\lambda
_{f}=\lambda +\delta $ with $\delta $ being a small quantity. Without loss
generality, we suppose that $H(\lambda )=H_{0}+\lambda H_{1}$. Since the
initial state is taken as the ground state of $H(\lambda )$, i.e., $\Psi
(0)=\psi _{0}(\lambda )$, we have
\begin{equation}
\overline{\mathcal{L}}_{\delta }=\sum_{n}\left\vert \langle \psi _{n}(%
\mathbf{\lambda +\delta })|\psi _{0}(\mathbf{\lambda })\rangle \right\vert
^{4} .  \label{L-delta}
\end{equation}

Expanding the wave function $|\psi _{0}(\mathbf{\lambda }+\delta )\rangle $
in the basis of eigenstates corresponding to the parameter $\lambda $, to
the first order of $\delta $, we get {\small
\begin{equation}
|\psi _{n}(\lambda +\delta )\rangle =c_{n}\left( |\psi _{n}(\mathbf{\lambda }%
)\rangle +\delta \sum_{m\neq n}\frac{H_{mn}|\psi _{m}(\mathbf{\lambda }%
)\rangle }{E_{n}(\lambda )-E_{m}(\lambda )}\right) ,  \label{Psi_n}
\end{equation}%
} where $c_{n}=\left\{1+\delta ^{2}\sum_{m\neq n}|H_{mn}|^{2}/\left[
E_{n}(\lambda )-E_{m}(\lambda )\right]^2 \right\}^{-1/2}$
are the normalization constants and $H_{mn}=\langle \psi _{m}(\mathbf{%
\lambda })|H_{1}|\psi _{n}(\mathbf{\lambda })\rangle $. Substituting the
conjugation of Eq.(\ref{Psi_n}) into Eq.(\ref{L-delta}) and expanding $%
\overline{\mathcal{L}}_{\delta } $ to the second order of $\delta $, we have%
\begin{equation}
\overline{\mathcal{L}}_{\delta }=1-2\delta ^{2}\sum_{m\neq 0}\frac{%
\left\vert H_{m0}\right\vert ^{2}}{\left[ E_{0}(\lambda )-E_{m}(\lambda )%
\right] ^{2}}.
\end{equation}%
Then, the term which defines the response of the $\overline{\mathcal{L}}%
_{\delta }$ to a small change in $\delta $ can be obtained as%
\begin{eqnarray}
\chi _{\delta } =-\frac{\partial ^{2}\overline{\mathcal{L}}_{\delta }}{%
\partial \delta ^{2}} =\sum_{m\neq 0} \frac{4\left\vert H_{m0}\right\vert ^{2}}{\left[
E_{0}(\lambda )-E_{m}(\lambda )\right] ^{2}}\, .  \label{LLS}
\end{eqnarray}

Now we explore the relation between $\chi _{\delta }$ and the fidelity
susceptibility. We notice that the ground state fidelity is defined as the
overlap of wavefunctions with driving parameter $\lambda $ and $\lambda
+\delta $ \cite{PZ2006PRE}, i.e.,
\begin{equation}
\mathcal{F}=\left\vert \langle \psi _{0}(\mathbf{\lambda }+\delta )|\psi
_{0}(\mathbf{\lambda })\rangle \right\vert \, .  \label{Fidelity}
\end{equation}
Substituting the conjugation of Eq.(\ref{Psi_n}) into Eq.(\ref{Fidelity}),
we have%
\begin{equation}
\mathcal{F}=\left( 1+\delta ^{2}\sum_{m\neq 0}\frac{\left\vert
H_{m0}\right\vert ^{2}}{\left[ E_{0}(\lambda )-E_{m}(\lambda )\right] ^{2}}%
\right) ^{-1/2},
\end{equation}%
and the fidelity susceptibility is given by \cite{GuSJ,Zanardi}
\begin{eqnarray}
\mathcal{\chi }_{\mathcal{F}} =-\frac{\partial ^{2}\mathcal{F}}{\partial
\delta ^{2}} =\sum_{m\neq 0} \frac{\left\vert H_{m0}\right\vert ^{2}}{\left[ E_{0}(\lambda
)-E_{m}(\lambda )\right] ^{2}}\, .
\end{eqnarray}%
The connection of fidelity susceptibility and the Berry curvature was discussed in the reference \cite{Zanardi}.
A review article for the role of fidelity and fidelity susceptibility  in the characterization of static QPTs can be found in the reference \cite{GuSJ2}.
In comparison with Eq.(\ref{LLS}), it is straightforward to find the
following relation:
\begin{equation}
\chi _{\delta }=4\mathcal{\chi }_{\mathcal{F}}\, .  \label{LL_F}
\end{equation}

\begin{figure}[tbph]
\includegraphics[width=9cm]{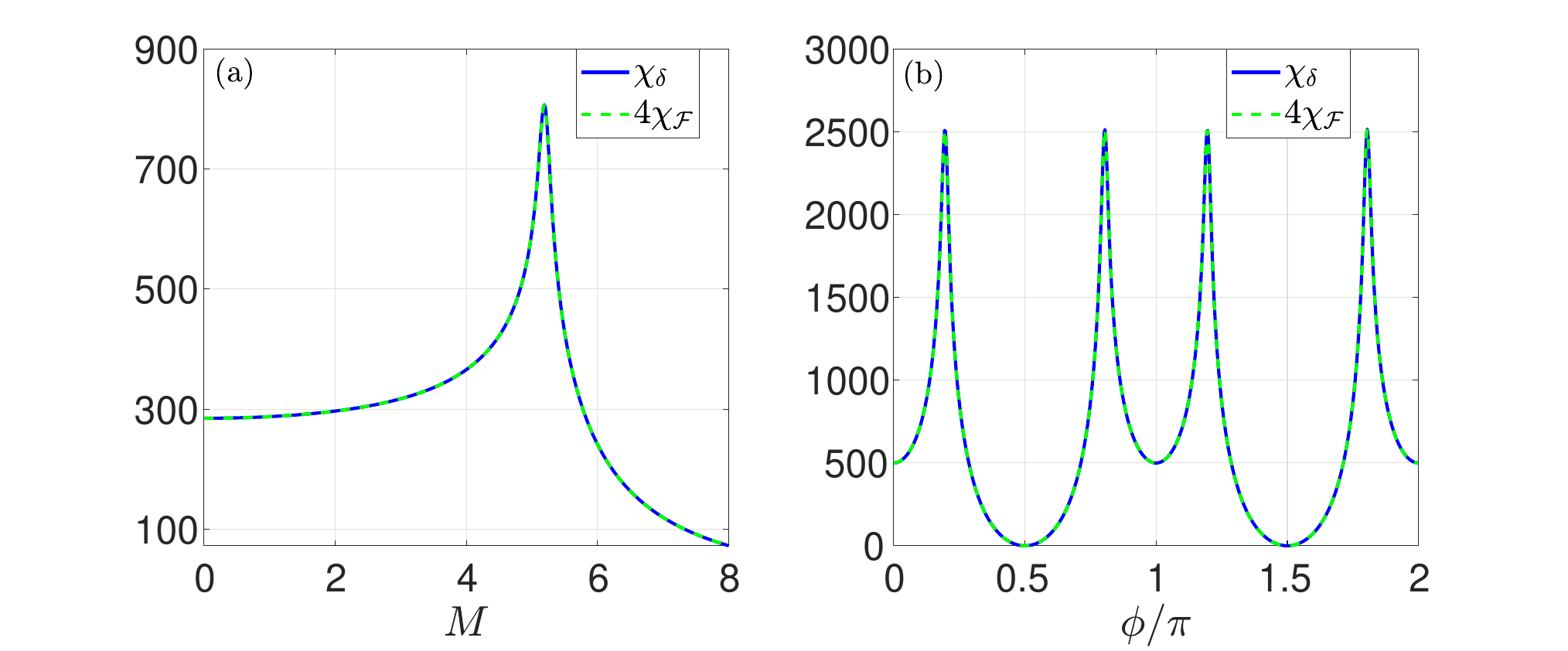}
\caption{(Color online) (a) $\protect\chi _{\protect\delta }$ and $4\mathcal{%
\protect\chi }_{\mathcal{F}}$ versus $M$. For the quench process, we take $%
M_i=M$ and $M_f=M+\protect\delta$. The parameters are $\protect\delta %
=10^{-5}$ and $\protect\phi =\protect\pi /2$. (b) $\protect\chi _{\protect%
\delta }$ and $4\mathcal{\ \protect\chi }_{\mathcal{F}}$ versus $\protect%
\phi $. Here we take $\protect\phi_i=\protect\phi$ and $\protect\phi_f=%
\protect\phi+\protect\delta$. The parameters are $\protect\delta =10^{-5}$
and $M=3$. }
\label{LLFS}
\end{figure}

In Fig.\ref{LLFS} (a) and (b) we numerically illustrate $\chi _{\delta }$
and $4\mathcal{\chi }_{\mathcal{F}}$ versus $M$ and $\phi$ for the Haldane
model, respectively. It is found that the two curves are identical,
consistent with the analytical relation given by Eq.(\ref{LL_F}). It is well
known that the fidelity susceptibility is divergent at the phase transition
point \cite{GuSJ,SC2008PRA,Zanardi}. The relation between $\chi_{\delta}$
and $\mathcal{\chi }_{\mathcal{F}}$ suggests the existence of divergence in $%
\chi_{\delta }$ around the phase transition point.

\section{Summary}

In summary, we have studied the long time average of LE for the sudden
quench processes in various quantum systems, including the AA model, quantum
Ising model and Haldane model, and shown that the long time average of LE $%
\overline{\mathcal{L}}(\lambda _{f})$ or its rate function $\eta (\lambda
_{f})$ exhibits nonanalytic behavior when the quench parameter crosses the
phase transition points. For the AA model and quantum Ising model, we
demonstrated that as quench parameter varies across a phase transition
point, the long time average of LE or its rate function has an obviously
sudden change around the transition point. For the Haldane model, the
nonanalyticity of the rate function at the phase transition point is not so
obvious. But we found the quantity $\chi _{\lambda _{f}}$ which is
proportional to the second derivative of rate function exhibits a divergent
peak as the quench parameter crosses the phase transition points.
Considering the limiting case that the pre-quench and post-quench parameters
are very close, we analytically proved that $\chi _{\delta }$ is
proportional to the fidelity susceptibility as $\delta \rightarrow 0$. The
connection with fidelity susceptibility suggest that the long time average
of LE and its rate function can be used to signal nonequilibrium QPTs in
more general systems.

\begin{acknowledgments}
The work is supported by NSFC under Grants 11974413 and 11425419 and the National Key
Research and Development Program of China (2016YFA0300600 and
2016YFA0302104).
\end{acknowledgments}

\end{document}